\documentclass[aps,prl,paper,reprint,amsmath,amssymb,superscriptaddress,floatfix,showpacs,showkeys,subeqn,footinbib]{revtex4-1} 
\usepackage{amsmath, amsfonts, graphics, graphicx, hyperref,url,array, multirow, subfigure, rotating}
\usepackage{todonotes}
\usepackage{color}

\newcommand{\beq}{\begin{equation}}
\newcommand{\eeq}{\end{equation}}

\begin{document}

\title{Random close packing fraction of bidisperse discs: theoretical derivation}
\author{Raphael Blumenfeld}
\affiliation{Gonville \& Caius College, University of Cambridge, Trinity St., Cambridge CB2 1TA, UK}

\date{\today}

\begin{abstract}
Predicting theoretically the highest density, which a disordered packing of discs can achieve, has been a long-standing unresolved problem. 
Such predictions are hindered by two difficulties - the dependence of the density on the packing procedure and ensuring disorder. 
A theory that overcomes these difficulties has been developed recently for mono-disperse disc packing~\cite{Bl21}. However, to minimise order, experiments and numerical simulations often use two-size discs and a prediction of the highest possible packing fraction, $\phi_{RCP}$, for these  packings is arguably more useful. This problem is more complex because in such packings, $\phi_{RCP}$ is not a number but a function of the sizes ratio, $D$, and concentrations, $p$, of the disc types. 
A disorder-guaranteeing theory is formulated here to derive $\phi_{RCP}(p,D)$ under some assumptions, using the concept of the cell order distribution. Exact upper and lower bounds on the densest disordered packing fraction are also derived.   

\end{abstract}

\maketitle


\noindent\textbf{Introduction}

Packing objects densely is a goal of practical importance in technology and more generally in human society. The quest to predict the packing fraction of the densest possible packing of a given number of objects focused much research. Special attention has been given to a canonical version of the problem - predicting the highest packing fraction of disordered packings of spheres in three dimensions  ($d=3$) and discs in two ($d=2$), named the random close packing (RCP) problem. This problem is relevant to many technological applications~\cite{Toetal10}, to liquid-solid phase transitions~\cite{Bernal}, to information packing~\cite{IT1,IT2,IT3}, to process and nuclear engineering~\cite{Proc,Nucl}, as well as to the understanding and modelling of many natural phenomena. 
A successful theoretical approach to solving the canonical problem has the potential to pave the way to predicting and facilitating packing of more general objects. 

Many experimental and numerical studies put the highest packing fractions of same-size (monodisperse) spheres and discs, $\phi_{RCP}$, inside narrow ranges of values, e.g., $0.82-0.86$ for monodisoperse discs in $d=2$~\cite{Kaetal71,QuTa74,BiTr84,Hietal90,Oheetal01}. However, such studies use packings generated physically or numerically and, as such, are part of a collection of trial-and-error cases because they do not rule out a yet-untried process that could generate a denser state. 
Two main obstacles stand in the way of a theoretical resolution of the problem. One is that there are many ways to generate a packing of any given set of objects, each giving rise to a different packing fraction. As the parameters of each process can be varied continuously, this means that the problem exists in an infinite-dimensional parameter space~\cite{Bl21}, which no trial-and-error approach can exhaustively explore. 
Another obstacle, especially prominent in the canonical problems, is ensuring disorder - under most packing processes, monodisperse spheres and discs tend to crystalize into ordered states that are denser than disordered ones, e.g., hexagonal order in $d=3$ and trigonal in $d=2$. Thus, deriving $\phi_{RCP}$ theoretically requires an  {\it a-priori} criterion that ensures disorder. While several disorder criteria have been proposed in the literature, e.g., in~\cite{Toetal00}, those are difficult to implement because they are often a-posteriori in the sense that they require testing disorder in already-existing packings. Such criteria must then follow the generattion of packings, which give rises unavoidably to process-dependence and running against the infinite-parameter-space difficulty. 
Both these obstacles can be circumvented in $d=2$ packings of same-size discs by using the cell order distribution (COD), as discussed below. 

While the prediction of $\phi_{RCP}$ appears reasonably resolved for monodisperse discs in $d=2$~\cite{Bl21}, to minimise crystallisation, many simulations and experiments are carried out on packings of bidisperse discs, e.g.,~\cite{ViBo72,Bietal86,OkOd04,Xuetal05,Koetal16,Br25}. Therefore, a prediction of $\phi_{RCP}$ in such systems is, arguably, more useful. The aim here is to derive the highest possible packing faction, $\phi_{RCP}$, of any disordered packing of bidisperse discs in $d=2$.  
This problem is more complex than the monodisperse one because, in addition to the aforementioned difficulties of the infinite-dimensional parameter space and minimising order, $\phi_{RCP}$ is not a number in bidisperse  packings but rather a function of the disc sizes ratio, $D$, and the concentration of the smaller discs, $p$. 
The aim here is to generalise the approach developed in~\cite{Bl21} to address this more complex problem. 

Without loss of generality, it is sufficient to consider the size ratio of the two disc types, $D$, regarding the diameter of the smaller discs to be a unit length. 
The following analysis is limited to disc size ratios $1<D<D_{max}=\sqrt{3}/(2-\sqrt{3})$, with the value of $D_{max}$ ensuring that no `rattlers' can nestle within enclosures of three large discs. 
The concentration of the small discs is a relevant parameter, denoted by $p$ and the packings are assumed isotropic and very large. The latter assumption ensures that all possible structural configurations are realised within the packing and boundary corrections can be neglected. 

This paper's structure is the following. The COD is defined in the next section and its advantages are briefly described. An a-priori disorder criterion is then proposed, which makes it possible to identify the range of $p(D)$, within which bidisperse packings are assured to be disordered. In the following two sections, exact upper and lower bounds on $\phi_{RCP}(p,D)$ are derived. Confining the packings to the ranges that ensure disorder, the value of $\phi_{RCP}$ is then derived as a function of $p$ and $D$. The paper is concluded with a summary of the results and a discussion of potential uses and extensions. \\

\noindent\textbf{The COD}

The COD is defined for general planar $N$-disc systems as follows~\cite{MaBl14,Waetal20}. One connects the centres of each disc with those of the nearest neighbours it contacts, which generates a graph whose nodes are the disc centers and whose edges are the lines connecting them. The smallest voids enclosed by the edges are the 'cells'. The number of edges surrounding a cell is defined as its order and the distribution of this number is the COD. The fraction of cells of order $k$ (henceforth, $k$-cells) out of all existing $N_c$ cells is denoted $Q_k$. The packings considered here contain $N_c\gg1$ cells and $N\gg1$ discs, a fraction $p$ of which of unit diameter and the rest are of diameter $D$.  

As any packing process must generate some COD, one can use this distribution to classify and practically parameterise packings. Using the COD then circumvents the difficulty of the infinite process parameter space~\cite{Bl21}. As the COD can be related directly to the density in the large $N_c$ limit, as is shown explicitly next, it serves as a very useful tool to address the RCP problem. Given $p$, $D$, and an a-priori disorder criterion, we can translate this problem to the problem of finding the COD that provides the highest density. In turn, this allows us to predict $\phi_{RCP}$. Clearly, different packing procedures can give rise to packings with the same COD. While the dynamic and mechanical properties of those packings might differ, e.g., if they have different friction coefficients, the would nevertheless have the same packing fraction, as long as the packings are  isotropic and $N_c\to\infty$.

Given the COD, let $\bar{S}_{k}$ be the mean area of the $k$-cells over all their configurations. Then, the total area of the packing per cell, irrespective of its order, is
 \begin{align}
S_{pack} = N_c\sum_{k=3}^\infty Q_k\bar{S}_{k}  \ ,
\label{Spack}
\end{align}
The packing fraction is the ratio of the total area of the discs contained within the cell polygons per cell, $S_{disc}$, and $S_{pack}$. Although the discs at the periphery of the packings stick outside of the cell polygons, this introduces a correction that vanishes as $1/\sqrt{N_c}$ as $N_c\to\infty$. 

\begin{figure}[h]
\includegraphics[width=.8\linewidth]{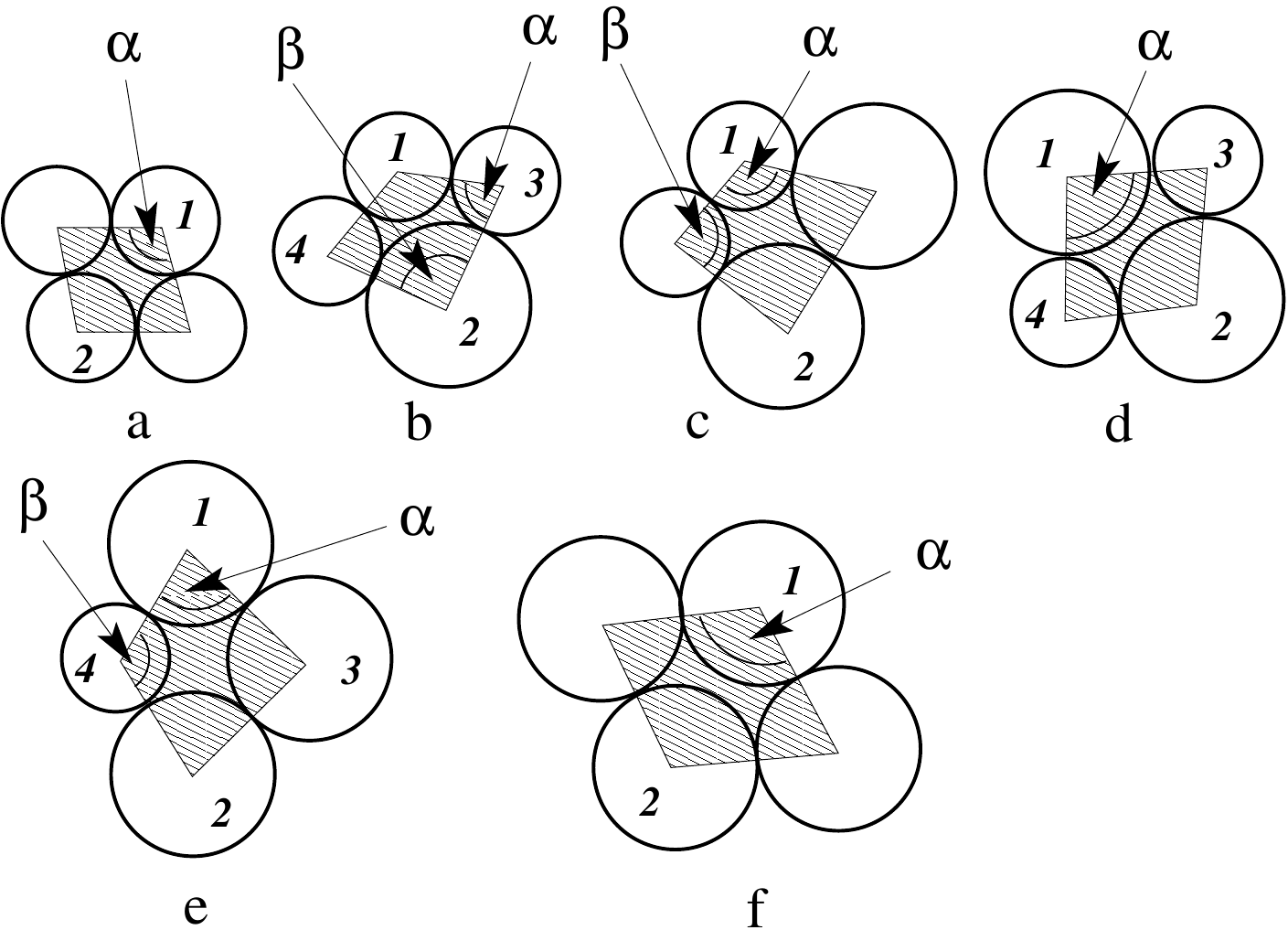}
\caption{The six possible 4-cell configurations and their areas, $S_{4i}$, $i=a,b,c,d,e,f$ (shaded). The areas the discs occupy within the shaded cell areas, $S^{disc}_{4i}$, depend on the internal angles of the quadrilateral cells, e.g., $\alpha$ and $\beta$.}
\label{4combs}
\end{figure}

To calculate $S_{disc}$, note that the area that a disc occupies inside a cell polygon depends on the internal polygon angles, e.g., angles $\alpha$ and $\beta$ in the $4$-cell configurations shown in Fig.~\ref{4combs}. The internal angles of a $k$-cell sum up to $(k-2)\pi$ and, averaging over the disc sizes in all the $k$-cell configurations, the total area occupied by the discs within the $k$-cells is $N_c(k-2)\pi\left[p+(1-p)D^2\right]/8$. 
Averaging over the COD, the packing fraction is then,
\begin{align}
\phi &= \frac{\sum_{k=3}^\infty Q_k \frac{(k-2)\pi}{8}\left[p+(1-p)D^2\right]}{S_{pack}}N_c \notag \\
&= \frac{\pi\left[p+(1-p)D^2\right]\left(\bar{k}-2\right)}{8S_{pack}} N_c \ ,
\label{Phi}
\end{align}
where $\bar{k}=\sum_{k=3}^\infty kQ_k$ is the first moment of the COD. 
This expression is valid for any isotropic packing realisation when $N_c\gg1$, which establishes that all packings with the same $p$, $D$, and COD have the same packing fraction. 

Relation (\ref{Phi}) makes it clear that, given $p$ and $1<D<D_{max}$, the highest possible packing fraction of any bidisperse packing is obtained when $\bar{k}$ is the lowest. This is as expected because increasing the fraction of low-order cells increases the mean number of contacts per disc, which increases monotonically with the packing fraction. The inter-dependence between $\bar{k}$ and the packing fraction is discussed in more detail in the supplemental material~\cite{SM}. 
Thus, the random packing problem has been reduced to the problem of finding the COD with the highest possible fraction of low-order cells while keeping the packing disordered. The next question is then how to minimise $\bar{k}$ while satisfying the disorder constraint without the need to measure it post-packing. This is the subject of the next section - development of an a-priori disorder criterion that limits the occurrence probability of large ordered clusters. \\

\noindent\textbf{Ensuring disorder}

Several disorder measures and criteria have been suggested in the literature, e.g., in~\cite{ZePr27,Stetal83,Jietal11,Moetal24}. However, those are mainly geared to measure the disorder of packings that had already been generated. As each packing's structure depends on the generating procedures, using any of those post-packing measures is unavoidably specific to the inspected packing, encountering the infinite-parameter-space difficulty. 
As $\phi_{RCP}$ must be the highest of all possible packings, it has to be procedure-independent and the only way to keep the derivation general is by ensuring disorder without the need to generate any packing at all. Such an a-priori disorder criterion is proposed in this section. 

\begin{figure}[h]
\includegraphics[width=.4\textwidth]{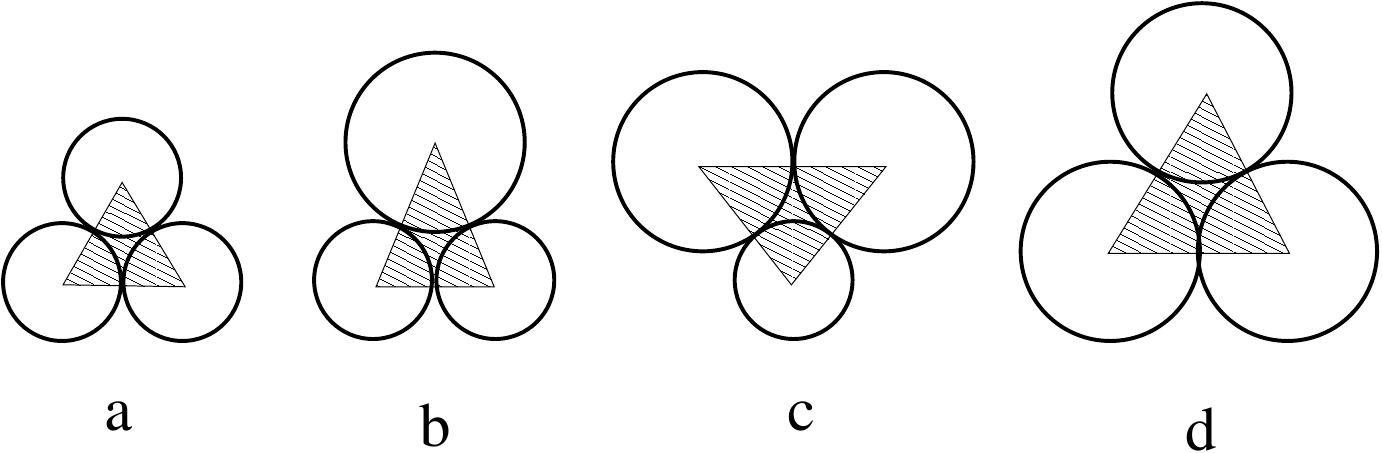}
\caption{The four possible $3$-cell configurations and their areas, $S_{3i}$, $i=a,b,c,d$ (shaded). The areas the discs occupy within the shaded cell areas are, respectively, $S^{disc}_{3i}$.}
\label{fig:3combs}
\end{figure}

In bidisperse disc systems, $3$-cells can occur in four configurations, whose triangular areas, $S_{3a}$, $S_{3b}$, $S_{3c}$, and $S_{3d}$, are shown shaded in Fig.~\ref{fig:3combs}. Order emerges in the form of large clusters of trigonal lattices of $3$-cells, consisting of tightly packed identical discs of either configuration $a$ or $d$. Large such clusters correspond to high fractions $Q_3$ and they are inescapable when $p\to0$ or $1$. The rationale behind the disorder criterion proposed here is, therefore, to suppress the occurrence of such clusters by limiting $Q_3$. The immediate objective of the criterion developed here is to identify the allowed range of $p$, within which the occurrence probability of such a cluster is very low.  

The criterion, proposed here, is bassed on limiting the neighbouring of identical equilateral cells, namely, $3$-cells of either configuration $a$ or $d$ in Fig.~\ref{fig:3combs}. It consists of requiring that the probability of such a cell having more than one identical cells is smaller than $1/3$. To implement this criterion, lwe need to know the occurrence probabilities of all four $3$-cell configurations. In isolation, these would be, respectively,
\begin{align} 
&\left(p_{3a},p_{3b},p_{3c},p_{3d}\right) = \notag \\
&\left[p^3,3p^2(1-p),3p(1-p)^2,(1-p)^3\right]Q_3 \ .
\label{OccProb3}
\end{align} 
However, when such cells are adjacent these probabilities are not independent because the neighbours share two discs. I conjecture that this inter-dependence changes the occurrence probabilities in a manner that can be parameterised using the same form as in (\ref{OccProb3}), but with a modified probability $0\leq u\neq p\leq1$:
\begin{align}
&\left(p_{3a},p_{3b},p_{3c},p_{3d}\right) = \notag \\
&\left[u^3,3u^2(1-u),3u(1-u)^2,(1-u)^3\right]Q_3 \ .
\label{param}
\end{align} 
Strong support for this one-parameter conjecture will be provided below. For completing the theory, it is necessary to know the relation between $p$ and $u$ and this is derived later, eq.~(\ref{puRelation}).

Using the respective occurrence probabilities of configurations $a$ and $d$, $u^3Q_3$ and $(1-u)^3Q_3$, the probability that a cell of either configuration has more than one identical neighbour is 
\begin{align}
R_{a} = &u^3Q_3\left[ \left[u^3Q_3\right]^3 + 3\left(1 - u^3Q_3\right)\left(u^3Q_3\right)^2\right] \notag \\
R_{d} = &(1-u)^3Q_3\Big\{\left[(1-u)^3Q_3\right]^3 + \notag \\
&3\left[1 - (1-u)^3Q_3\right]\left[(1-u)^3Q_3\right]^2\Big\} \ .
\label{criteria}
\end{align}
Imposing that both $R_{a}<1/3$ and $R_{d}<1/3$, the respective solutions of (\ref{criteria}) are $u^3Q_3<A$ and $(1-u)^3Q_3<A$, with $A\equiv0.562$. The value of $A$, as well as the values of all  the numerical quantities appearing below, can be calculated to any desired accuracy and they are limited in this report to three decimal order only for brevity. 
Isolating the fraction of $3$-cells from the solutions of (\ref{criteria}), $Q_3$ must satisfy,
\beq
Q_3 < min\left\{\frac{A}{u^3},\frac{A}{\left(1-u\right)^3},1\right\} \ .
\label{Criteria1}
\eeq
In the supplemental material~\cite{SM}, I show that this criterion leads to an exponentially decaying probability of the ordered clusters with their size and that the probability that such a cluster is larger than five identical $3$-cells is lower than $1.56\times10^{-3}$ for any combination of $p$ and $D$.

\begin{figure}[tp]
\includegraphics[width=.8\linewidth]{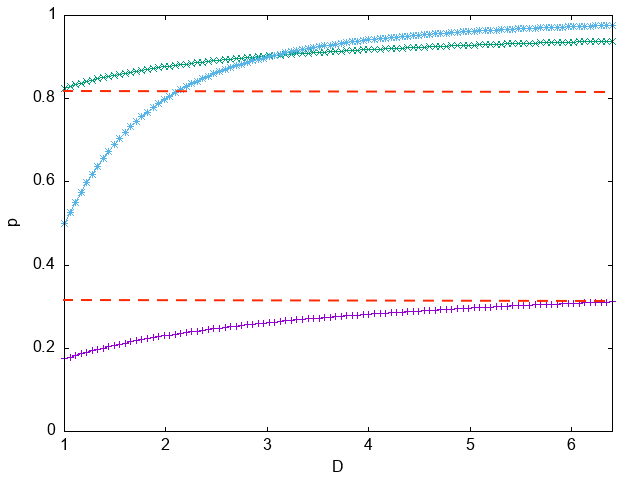}
\caption{For any choice of $D$, a 3-cells system would be disordered when $p$ is between $p_{min}(D)$ (purple curve) and $p_{max}(D)$ (green curve). Tthe lowest curve extends from $p_{min}(D=1)=0.175$ to $p_{min}(D=6.4)=0.312$ and the upper curve from $p_{max}(D=1)=0.825$ to $p_{max}(D=6.4)=0.939$. 
Disorder is then assured {\it for any value of $1<D<D_{max}$} within the region $p_{low}\equiv0.312<p<0.825\equiv p_{high}$, demarcated by the two horizontal dashed lines. Also shown in the figure is the curve corresponding to the common choice of both disc types occupying the same area (blue curve), in which $p$ depends on $D$. The upper bound drops considerably in these packings for $D\lesssim3$, limiting the range of $p$, for which disorder is assured.}
\label{pD}
\end{figure}

As eqs. (\ref{Criteria1}) show, $Q_{3max}$ varies with $u$, but $Q_3$ is also bounded by $1$. Using this information in (\ref{Criteria1}) provides the range of $u$, within which the packing is always disordered: $A^{1/3}=0.825 > u > 1-A^{1/3}=0.175$. Using the relation $p(u)$, shown in eq. (\ref{puRelation}) below, this range can be expressed in terms of $p$ and $D$ and the objective of determining the allowed range of $p$ for any value of $D$ has been achieved. This range is plotted in Fig.~\ref{pD}. In particular, this calculation shows that disorder is assured {\it for any value of $1<D<D_{max}$} when $0.312<p<0.825$, indicated by the dashed horizontal lines in the figure. 

This criterion holds for any bidisperse packing and, in particular, for two frequently used combinations of $p$ and $D$ in experiments and simulations. One is a ratio $D=\sqrt{2}$, for which the above calculation yields that disorder is assured when $0.203<p<0.852$. The other is that the two disc types occupy the same area, $p/[p+(1-p)D^2]=0.5$. This makes $p$ dependent on $D$, $p=D^2/(D^2+1)$ and, using the above relations, it is possible to calculate the disorder range for such packings. This range is also plotted in Fig.~\ref{pD} and it shows that this practice runs a higher risk of generating order for $D \lesssim3$. \\

As discussed above and evidenced by eq. (\ref{Phi}), the packing fraction increases as $\bar{k}$ decreases. Additionally, several arguments in the supplemental material~\cite{SM}) substantiate that packings density increases as the fraction of low-order cells increases. However, packings containing only $3$-cels may not be disordered, especially when $p$ approaches $0$ or $1$. Combining then this observation with the disorder criterion (\ref{Criteria1}), the value of $\phi_{RCP}(p,D)$ is expected to be attained for disordered packings that contain only $3$- and $4$-cells. This is also consistent with relation (\ref{Criteria1}) that shows that $\phi(p,D)$ is a monotonically increasing function of $Q_{3}$. 
In $3$- and $4$-cell packings, relation (\ref{Phi}) reduces to  
\beq
\phi = \frac{Q_{3}\bar{S}^{disc}_3 + \left(1-Q_{3}\right)\bar{S}^{disc}_4}{Q_{3}\bar{S}_3 + \left(1-Q_{3}\right)\bar{S}_4} \ ,
\label{Phi4}
\eeq
in which $\bar{S}^{disc}_3$ is the mean discs area within $3$-cells, averaged over the four $3$-cell configurations and  $\bar{S}^{disc}_4$ is the mean discs area within $4$-cells, averaged over the six $4$-cell configurations shown in Fig.~\ref{4combs}. \\


\noindent\textbf{Exact bounds on $\pmb{\phi_{RCP}}$} 

\noindent\underline{A. An upper bound}: \\
Before deriving $\phi_{RCP}(p,D)$, it is instructive to determine exact upper and lower bounds on it. As discussed above and shown in detail in the supplemental material~\cite{SM}, the packings' density increases with increasing fraction of low-order cells. Thus, $\phi_{RCP}$ is expected to be the packing fraction of packings containing only $3$- and $4$-cells and the higher the fraction of $3$-cells the denser the packing. Relation (\ref{Phi4}) makes it evident then that packings of only $3$-cell would be the densest, if they can be generated and made disordered. However, while ordered such packings have been shown to be constructed~\cite{Fe25}, it is unclear whether disordered are or even topologically possible. Nevertheless, the packing fraction of such a hypothetical bidisperse packing, $\phi_3$, provides an exact upper bound, $\phi_{RCP}(p,D)<\phi_3$. 
To derive $\phi_3$ we first need to know the areas of the triangular areas of the four $3$-cell configurations, $S_{3a}$, $S_{3b}$, $S_{3c}$, and $S_{3d}$, shown shaded in Fig.~\ref{fig:3combs}. These can be straightforwardly calculated for any combination of $p$ and $D$:  
\begin{align}
S_{3a} & = \frac{\sqrt{3}}{4} \quad ; \quad S_{3b} = \frac{\sqrt{D\left(D+2\right)}}{4} \notag \\
S_{3c} & =  \frac{D\sqrt{1+2D}}{4} \quad ; \quad S_{3d} = \frac{\sqrt{3}}{4}D^2\ .
\label{eq:3areas}
\end{align}
We also need the areas that the discs occupy within these triangles, $S^{disc}_{3a}$, $S^{disc}_{3b}$, $S^{disc}_{3c}$, and $S^{disc}_{3d}$, which can also be readily calculated for any $p$ and $D$: 
\begin{align}
S^{disc}_{3a} & = \frac{\pi}{8} \quad ; \quad S^{disc}_{3b} = \frac{\arccos{\frac{1}{D+1}}+D^2\arcsin{\frac{1}{D+1}}}{4}  \nonumber \\
S^{disc}_{3c} & =  \frac{D^2\arccos{\frac{D}{D+1}}+\arcsin{\frac{D}{D+1}}}{4} \quad ; \quad S^{disc}_{3d} =  \frac{\pi D^2}{8} \ . 
\label{Disc3Areas}
\end{align}

\begin{figure}[h]
\includegraphics[width=.9\linewidth]{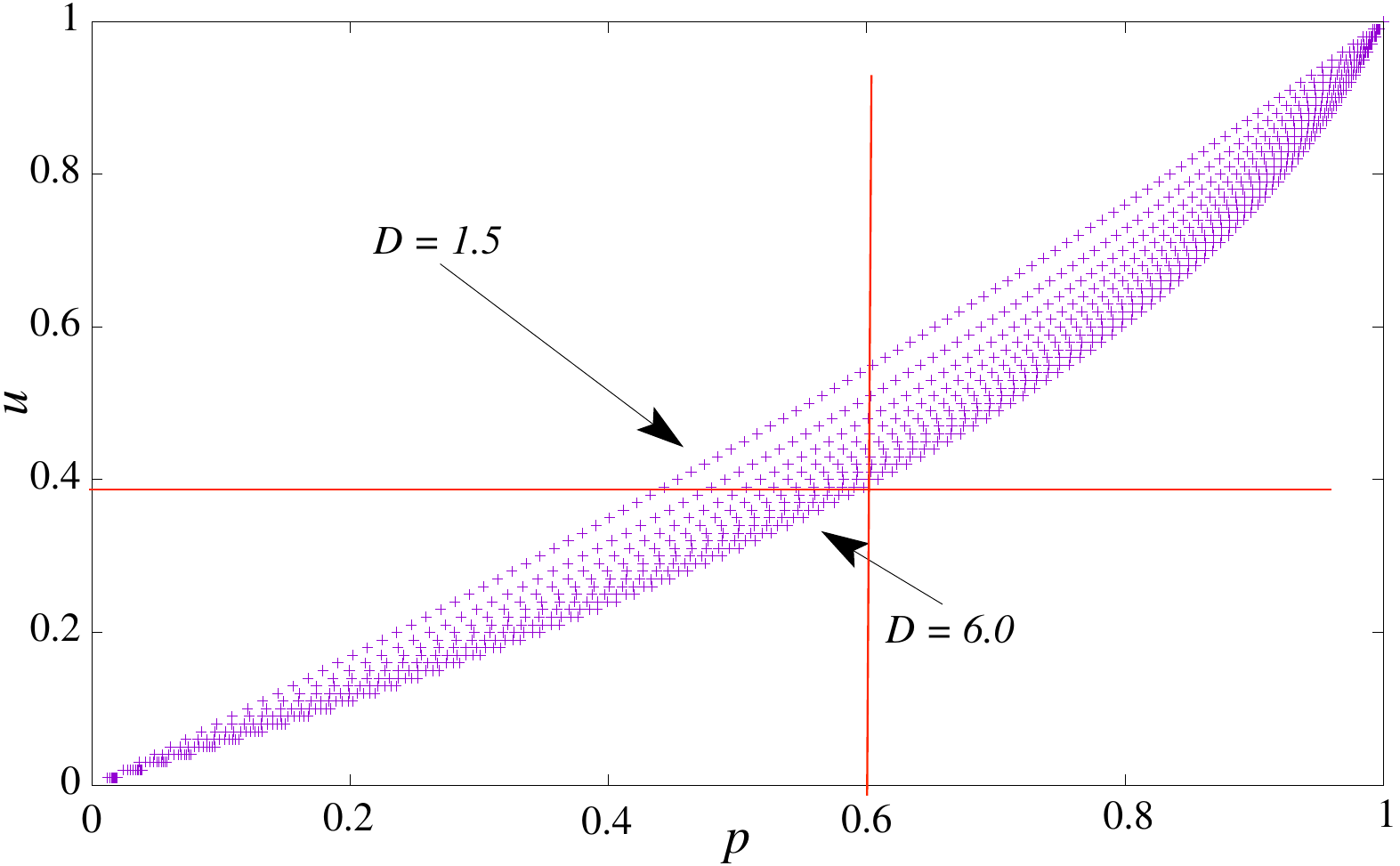}
\caption{The dependence of $u$ on $p$, described by eq.~(\ref{puRelation}), for several values of $D<D_{max}$. }
\label{fig:pu}
\end{figure}

Using then the occurrence probabilities of the four $3$-cell configurations in (\ref{param}), together with eqs. (\ref{eq:3areas}) and (\ref{Disc3Areas}), the mean total area, $\bar{S}_3=N_c s_{pack}$, is 
\begin{align}
\bar{S}_3 = & N_c\sum_{i=1}^4 p_{3i} S_{3i} = \frac{\sqrt{3}N_c}{4}\bigg[u^3 + (1-u)^3D^2 + \nonumber \\
& u(1-u)\sqrt{3}\left(\sqrt{D(D+2)}u + D\sqrt{2D+1}(1-u)\right) \bigg] \ .
\label{Total3area}
\end{align}
The mean disc areas enclosed in these $3$-cells is
\begin{align}
\bar{S}^{disc}_3 &= N_c\sum_{i} p_{3i} S^{disc}_{3i} = \frac{\pi N_c}{8}\Bigg[\left(u^3 + (1-u)^3D^2\right) + \nonumber \\
&\frac{6u(1-u)}{\pi}\bigg[u\left(\arccos{\frac{1}{D+1}}+D^2\arcsin{\frac{1}{D+1}}\right) + \notag \\
&v\left(D^2\arccos{\frac{D}{D+1}}+\arcsin{\frac{D}{D+1}}\right)\bigg]\Bigg] \ .
\label{TotalDisc3area}
\end{align}
The averages in (\ref{Total3area}) and (\ref{TotalDisc3area}) are over all the possible cell configurations in isotropic bidisperse packings as $N,N_c\to\infty$. Therefore, they hold true for all realisations of bidisperse $3$-cell packings. Note now that the total disc area is also expected to be $\bar{S}^{disc}_3=\pi\left[p+(1-p)D^2\right]N/4$ in this limit. Comparing this expression to (\ref{TotalDisc3area}) and noting that $3$-cell-only packings must satisfy Euler's topological relation $N_c=2N$, provides the relation between $p$ and $u$:
\begin{align}
p = u^3 &+ \frac{6u(1-u)}{\pi}\Bigg(u\arccos{\frac{1}{D+1}} \notag \\
&+ (1-u)\arcsin{\frac{D}{D+1}}\Bigg) + {\cal{O}}\left(\frac{1}{\sqrt{N_c}}\right) \ ,
\label{puRelation}
\end{align}
which is $D$-dependent, as expected. This relation is plotted in Fig.~\ref{fig:pu} for several values of $1.5\leq D\leq6.0$. 
In passing, it should be mentioned that relation (\ref{puRelation}) was used to test experimentally the conjecture that the occurrence probabilities of the $3$-cell configuration can be described by only one auxiliary probability $u$~\cite{SuZh26}. That test substantiated this conjecture, as described in more detail in the supplemental material~\cite{SM}. 

\begin{figure}[tp]
\includegraphics[width=.8\linewidth]{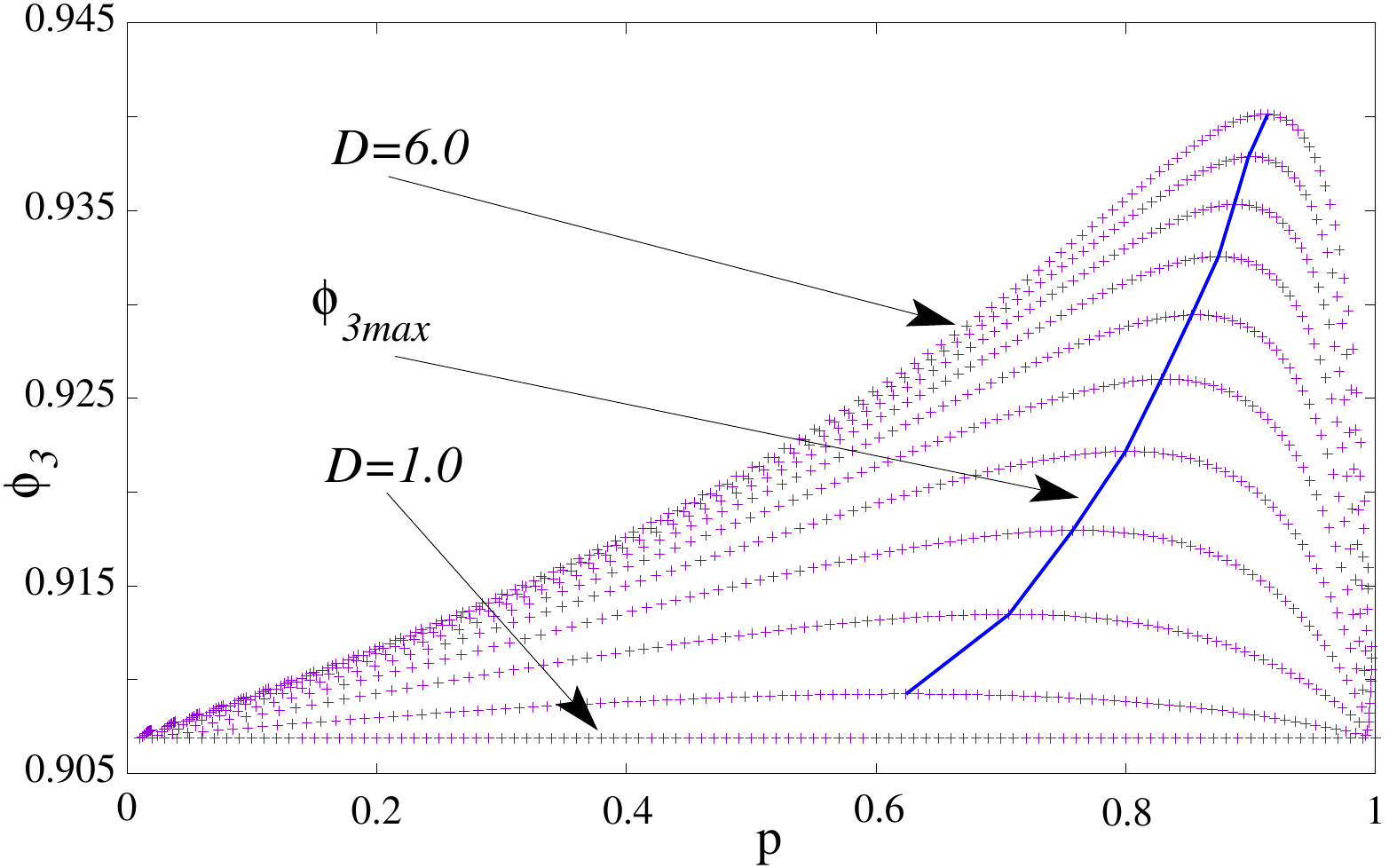}
\caption{The packing fraction, $\phi_3(p,D)$, for $1.0\leq D\leq6.0<D_{max}=\sqrt{3}/(2-\sqrt{3})$. 
The blue line follows the highest packing fraction for each mixture. $D=1.0$ recovers the fully trigonal crystal, $\phi_3=\pi/\left(2\sqrt{3}\right)$.} 
\label{fig:phi_u_D}
\end{figure}

\begin{figure}[h]
\includegraphics[width=.8\linewidth]{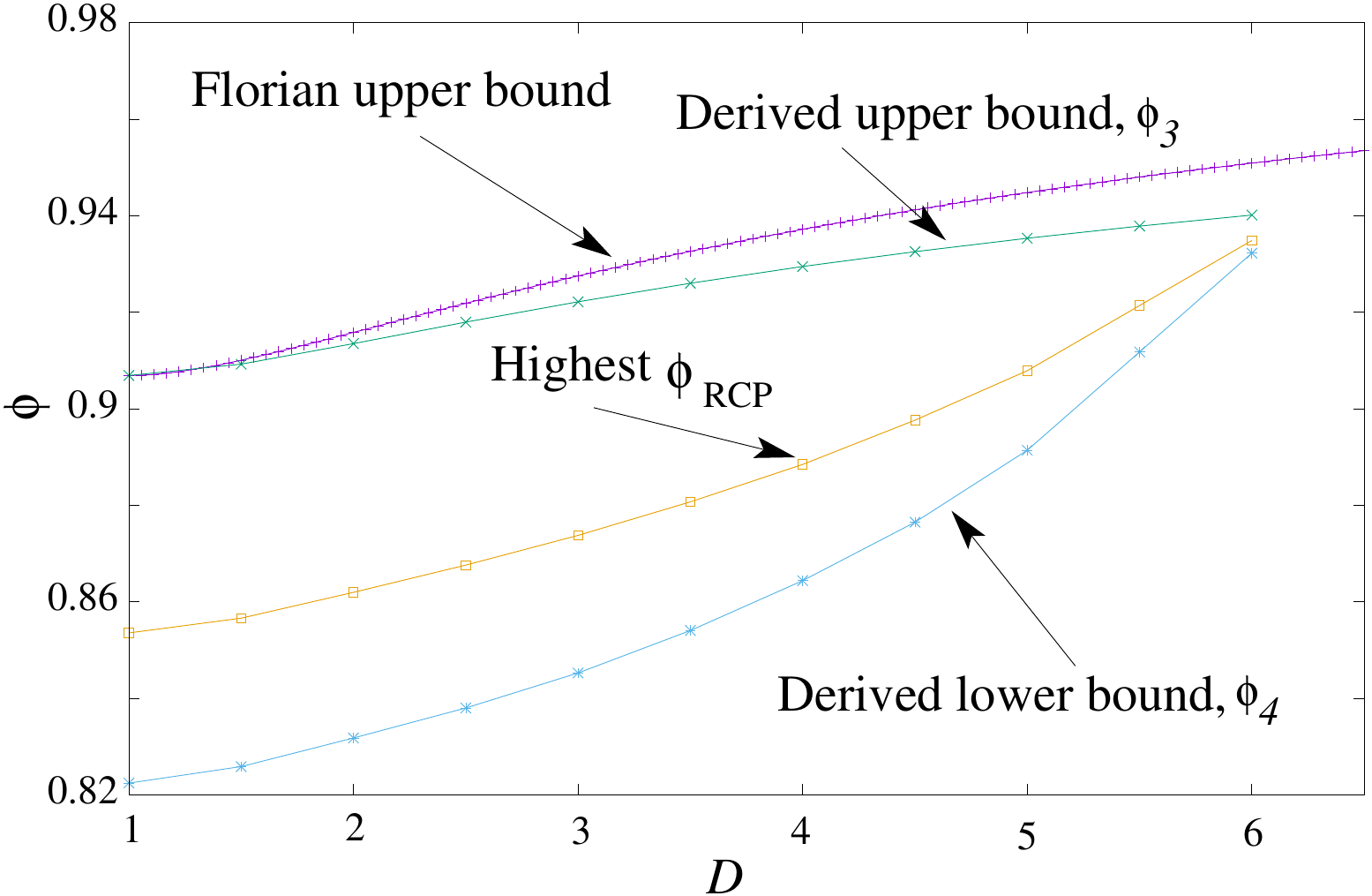}
\caption{The upper bound, $\phi_3$, improves on Florian bound~\cite{Fl60}, which is the specific packing fraction of the densest $3$-cell configuration, $b$ in Fig.~\ref{fig:3combs}.  Also plotted as functions of $D$ are the values derived in the main text for the highest value of $\phi_{RCP}(D)$, as well as the lower bounds on it, $\phi_4$.}
\label{PhiFPhi3PhiRCPPhi4}
\end{figure}

Using (\ref{Total3area}), (\ref{TotalDisc3area}), and (\ref{puRelation}), a plot of $\phi_3(p,D)\equiv\bar{S}^{disc}_3(p,D)/\bar{S}_3(p,D)$ is shown in Fig. \ref{fig:phi_u_D} as a function of $p$ for several values of $1<D<D_{max}$. Also shown is a line connecting the maxima of the curves, which provides the absolute maximum of $\phi_3$ for any specific $D$. Each such maximum is attained at a concentration $p_{max}(D)$ that depends on $D$. 
Each curve in Fig.~\ref{fig:phi_u_D} is an upper bound, $\phi_3(p,D)\geq\phi_{RCP}(p,D)$.  
It is noted that this upper bound improves on Florian's bound~\cite{Fl60}, which corresponds to an ordered packing of the configuration in Fig.~\ref{fig:3combs}b. The comparison between the two bounds can be seen in Fig.~\ref{PhiFPhi3PhiRCPPhi4}.
The function $\phi_3(p,D)$ has more use than just as a bound, it gives the packing fractions of all the {\it ordered} $3$-cell bidisperse disc packings, analysed and discussed in~\cite{Fe25}, with each of these packings corresponding to a specific combination of $p$, $D$, and $Q_3$. \\

\noindent\underline{B. A lower bound}: \\
As $\phi_{RCP}$ is expected to be derived for packings containing only $3$- and $4$-cells, it follows, as well as supported by the discussion in the supplemental material and relation (\ref{Phi4}), that any packing that contains only $4$-cells would be less dense. Therefore, the $4$-cell only packing fraction  is a lower bound, $\phi_{RCP}(p,D)>\phi_4$. 
This lower bound is the ratio of the mean cell areas, $\bar{S}_4$, and the mean areas occupied by the discs within them, $\bar{S}^{disc}_4$, with the averages carried out over the six $4$-cell configurations shown in Fig.~\ref{4combs}. These calculations are given in detail in the supplemental material~\cite{SM}. $\phi_4$ is plotted against $p$ for several values of $D$ between $D=1$ and $D=D_{max}$ in Fig.~\ref{fig:phi3phi4phi34}.  \\

\noindent\textbf{Derivation of $\pmb{\phi_{RCP}}$}

\begin{figure}[tp]
\includegraphics[width=.8\linewidth]{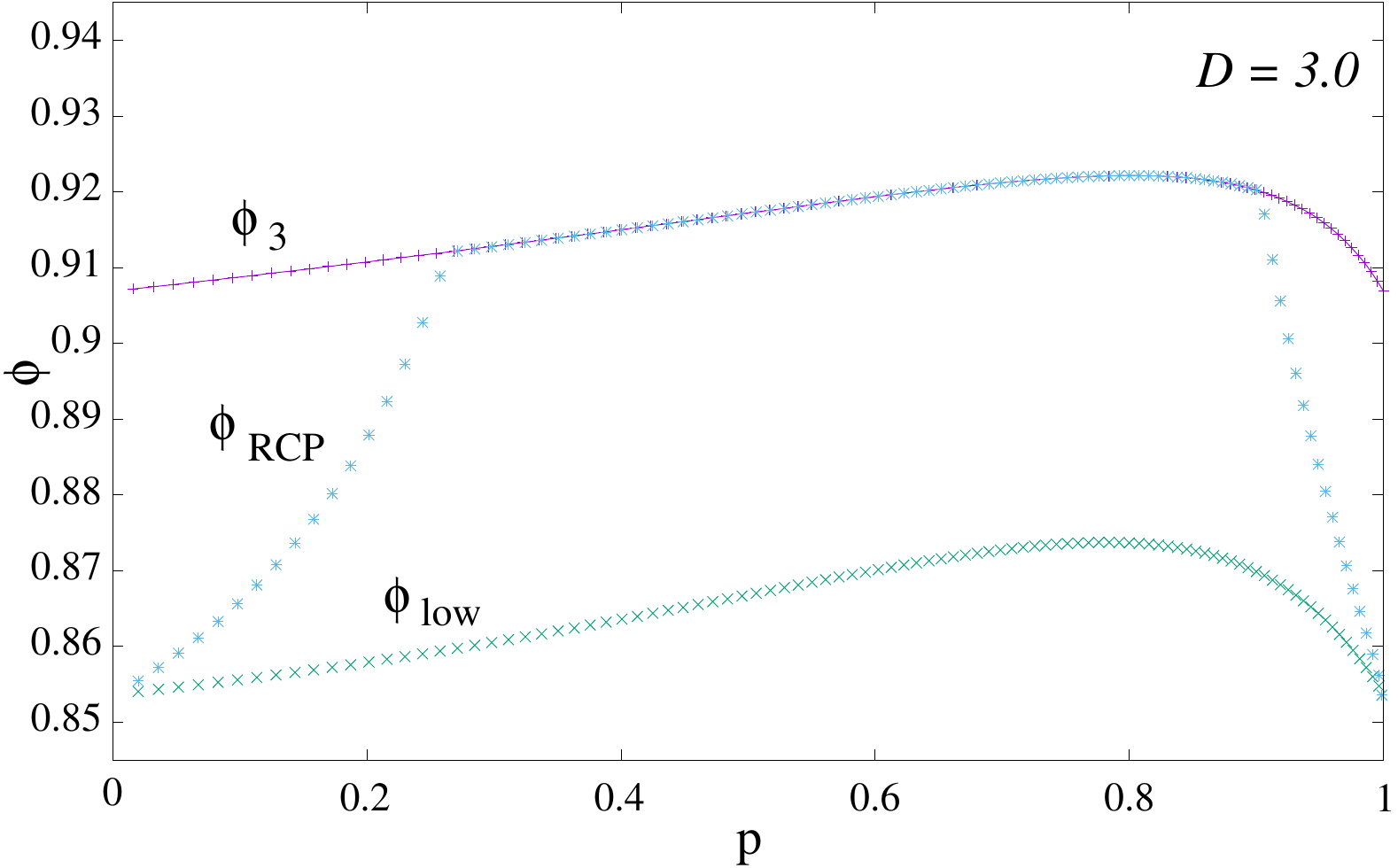}
\caption{A typical example of $\phi_{RCP}$ of bidisperse planar disc packing as a function of $p$ for $D=3.0$. It is bounded above by $\phi_3$ and below by $\phi_{low}$. }
\label{fig:D3phi3phi4phi34}
\end{figure}

\begin{figure}[tp]
\includegraphics[width=.8\linewidth]{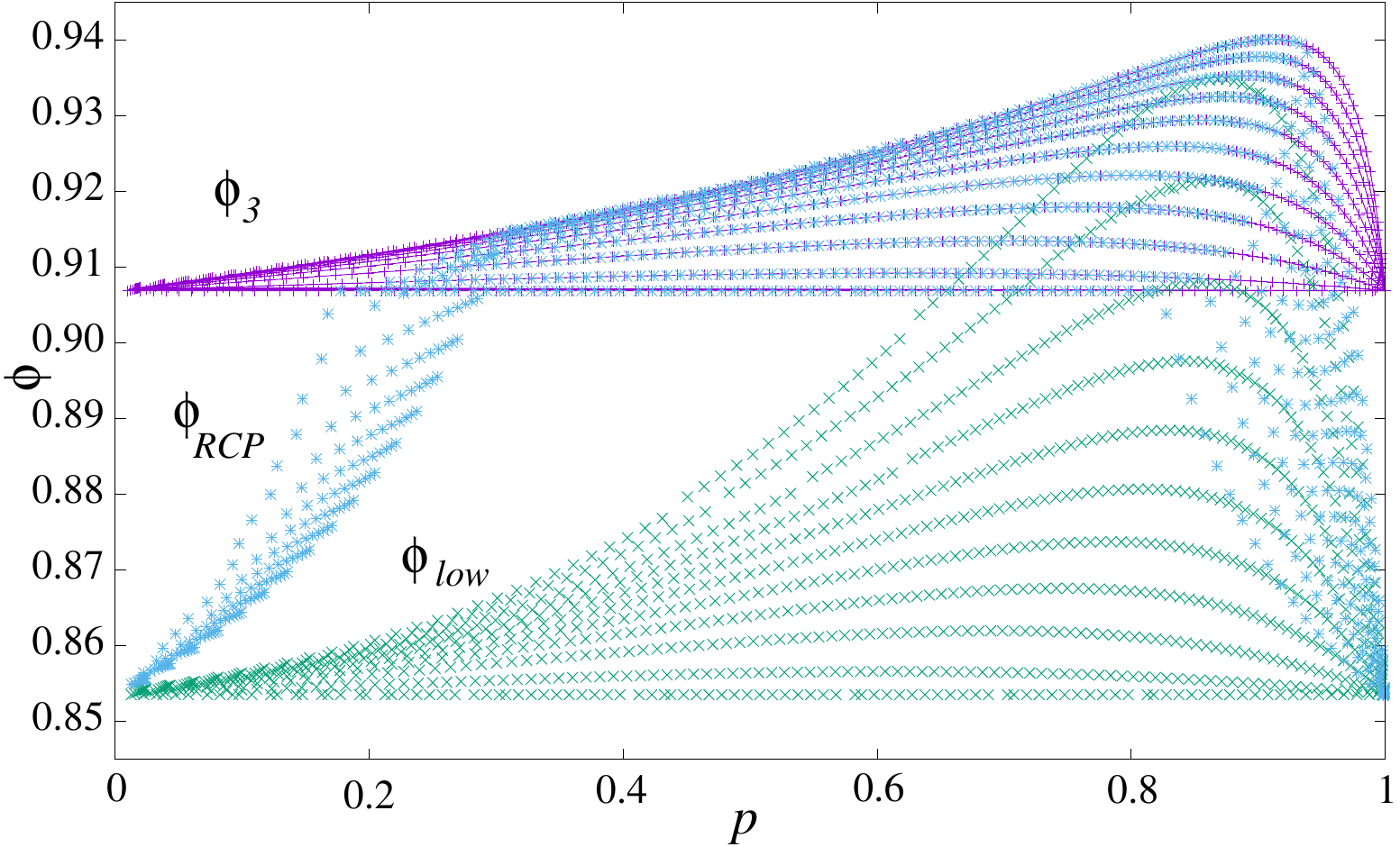}
\caption{As in Fig.~\ref{fig:D3phi3phi4phi34} for 11 size ratios $1.0\leq D\leq6.0$. For each value of $D$, the figure shows $\phi_{RCP}(p)$, as well as the upper and lower bound on it, derived here.}
\label{fig:phi3phi4phi34}
\end{figure}

As mentioned, $\phi_{RCP}$ corresponds to the packing fraction of a packing that contains only $3$- and $4$-cells at cell fractions that assure disorder, i.e. $Q_{3max}$ and $Q_4=1-Q_{3max}$. 
 The  packing's total area then consist only of $\bar{S}_3$ and $\bar{S}_4$, and that occupied by the discs of $\bar{S}^{disc}_3$ and $\bar{S}^{disc}_4$. Inspecting relation (\ref{Phi4}), $\phi_{RCP}(p,D)$ is achieved by minimising the fraction of $4$-cells while still inhibiting crystallisation, i.e., when $Q_3=Q_{3max}(p,D)$. To calculate it requires the averages $\bar{S}^{disc}_4(p,D)$ and $\bar{S}_4(p,D)$, which are derived in the supplemental material~\cite{SM}. All the quantities in relation (\ref{Phi4}) can be written in closed form, albeit very cumbersome, and $\phi_{RCP}(p,D)$ can be calculated to any desired precision. For the purpose of this report, its value was calculated numerically for a 100 values of $0\leq p\leq1$ and 11 values of $1\leq D\leq D_{max}$. A typical example for $D=3.0$, including the upper bound and lower bounds, is shown in Fig.~\ref{fig:D3phi3phi4phi34} and the aggregated curves for all $D$ are shown in Fig.~\ref{fig:phi3phi4phi34}.  
The plots show that, within a central region of $p$, $\phi_{RCP}$ coincides with the upper bound for any value of $D$. This is because, for these values of $p$ and $D$, $Q_{3max}=1$, namely, any random packing realisation would be disordered. Moreover, the highest possible packing fraction is within this region for any value of $D$.
This observation and the identification of the disorder-ensuring range of $p$ should aid planning compositions of very dense bidisperse packings in numerical and physical experiments. \\

\noindent\textbf{Summary and conclusions}

To conclude, a theory, based on the cell order distribution (COD), has been developed to solve analytically the random close packing problem for bidisperse packings of discs in $d=2$. The discs size ratio in this solution was limited in order to exclude rattlers in $3$-cells. 
Disorder-ensuring criteria have been proposed, which make it possible to determine, for any disc size ratio, the range of disc concentrations, $p$, within which the packing is disordered. The identification of this range for any choice of $D$ should assist in designing experiments and simulations that aim to study disordered such packings. In passing, it was shown that a common choice of bidisperse systems whose two disc types occupy the same area, increases the risk of crystallisation for $D\lesssim3$. 

Exact $p$- and $D$-dependent upper and lower bounds on $\phi_{RCP}$ have been derived, the former corresponding to a disordered packing of only $3$-cells and the latter to packings of only $4$-cells. To the best of this author's knowledge, these bounds improve on any bounds existing so far in the literature for such bidisperse packings.

Next, the exact value of the random close packing fraction, $\phi_{RCP}$, has been derived as a function of $p$ and $D$. It has been found that, for any size ratio, there is a range of concentrations, $p$, in which $\phi_{RCP}=\phi_3$ and that the highest possible packing fraction occurs within this range such that it always coincides with the upper bound. 
The use of the COD obviates the sensitivity to the packing process, which introduces an infinite parameter space that is impossible to explore fully by trial and error. Therefore, the results obtained here are valid in general, irrespective of the packing protocol and packing history. 

It should be commented that, in setting the disorder criterion, only limiting trigonal order has been considered. It could be argued that large clusters of some $4$-cells may also represent some kind of order because they represent deformed square lattices. However, it has been shown in the supplemental material~\cite{SM} that, while a disorder criterion similar to (\ref{Criteria1}) can also be derived for $Q_4$, it is not necessary. This is because packings that satisfy the criteria on the $3$-cell trigonal order, make the occurrence probability of deformed square lattices also exponentially small with cluster size. 

These results are useful to guide experiments and simulations, which rely on disordered disc packings. 
Firstly, identifying the exact packing fraction, at which the highest possible packing fraction is attained for any given value of $D$, can serve as a guide to future physical and numerical experiments in systems where high density is desired, even if the very densest state cannot be reached in practice. 
Secondly, measuring packing fractions that exceed the value predicted here would be an indication of extended crystalline domains eliminating the need to search the structure in detail to discover such order. 
Nevertheless, it should be emphasised $\phi_{RCP}(p,D)$ is the highest possible packing fraction that any bidisperse system can achieve and the derivation in this paper is independent of whether or not such a dense packings can be realised in practice. Nevertheless, a close examination, where possible, of existing simulations that aim to generate such packings~\cite{ViBo72,Bietal86,OkOd04,Xuetal05,Koetal16,Br25}, has found that $Q_3$ appears in none to be higher than the result for $Q_{3max}$ reported here. This is probably also why observed packing fractions of bidisperse disc packings are often lower than the theoretical values obtained here. This author hopes that the results obtained here could guide choices of $p$ and $D$, which can push experimental packings to higher densities than before. 

The method developed here is significant in its own right. 
It yields the packing fractions of ordered bidisperse packings that comprise of only $3$-cells, such as those discussed in~\cite{Fe25}.
It can also be used to analyze bidisperse disc packings with any COD, irrespective of its packing fraction. Such analyses require determining the statistics of cell orders higher than $4$ and, while this could be demanding to achieve analytically or in close form, there is no reason that such calculations cannot be done numerically.  

Another potentially exciting extension of the approach taken here is to analysis of packings that consist of discs of three sizes, $1$, $D_1$, and $D_2>D_1$, at respective concentrations $p$, $p_1$, and $p_2$. In these systems, the disorder criterion presented here should be applied to the three possible same-size $3$-cell configurations, namely, that none forms an extensive ordered cluster.  A two-parameter extension of (\ref{param}) for the $3$-cell configurational probabilities could be then attempted and tested. This extension is under way by this author and collaborators. 
Furthermore, there is no reason that such an extension should not be possible, at least in principle, to polydisperse disc packings with discrete disc size distributions, $1$, $D_1$, $D_2$, $\ldots$, $D_s$. However, the analytical calculations involved in such extensions are likely to becomes increasingly prohibitive with increasing polydispersity both because of the number of variables, which $\phi_{RCP}$ depends on, $2s$ for $s+1$ different disc sizes, and the number of possible configurations of cell structures, which increases combinatorically fast. It is possible, though, that such extensions can be achieved with computer-assisted calculations.


\begin{thebibliography}{99} 

\bibitem{Toetal10} S. Torquato, F.H. Stillinger, Rev. Mod. Phys. {\bf 82}, 2633 (2010) and references therein
\bibitem{Bernal} J.D. Bernal, Nature {\bf 185}, 68 (1960)
\bibitem{IT1} M. Parker, J. Ryan, Telecommunication Systems {\bf 2}, 185 (1994)
\bibitem{IT2} K. Park, S. Kang, S. Park, Management Science {\bf 42}, 1277 (1996)
\bibitem{IT3} A. Amiri, R. Barkhi, Euro. J. Oper. Res. {\bf 208} 37 (2011)
\bibitem{Proc} A. R. Brown, {\it Optimum packing and depletion: the computer in space - and resource-usage problems}, (Macdonald, NY, London, 1971)
\bibitem{Nucl} H. Suikkanen, J. Ritvanen, P. Jalali, R. Kyrki-Rajamäki, Nuclear Eng. Design {\bf 203}, 24 (2014)
\bibitem{Kaetal71} H.H. Kausch, D.G. Fesko, N.W. Tschoegl, J. Colloid Interface Sci. {\bf 37}, 603 (1971)
\bibitem{QuTa74} T. J. Quickenden and G. K. Tan, J. Colloid Interface Sci. {\bf 48}, 382 (1974)
\bibitem{BiTr84} D. Bideau, J.P. Troadec, J. Phys. C: Solid State Phys. {\bf 17}, L731 (1984)
\bibitem{Hietal90} E.L. Hinrichsen, J. Feder, T. J{\o}ssang, Phys. Rev. A {\bf 41}, 4199 (1990)
\bibitem{Oheetal01} C.S. O'Hern, S.A. Langer, A.J. Liu, S.R. Nagel, Phys. Rev. Lett. {\bf 86}, 111 (2001)
\bibitem{Bl21} R. Blumenfeld, Phys. Rev. Lett. {\bf 127}, 118002 (2021)
\bibitem{Toetal00} S. Torquato, T.M. Truskett, P.G.  Debenedetti, Phys. Rev. Lett. {\bf 84}, 2064 (2000)
\bibitem{ViBo72} W.M. Visscher, M. Bolsterli, Nature {\bf 239}. 504  (1972)
\bibitem{Bietal86} D. Bideau, A. Gervois, L. Oger, J. P. Troadec, J. Physique {\bf 47}, 1697 (1986) 
\bibitem{OkOd04} T. Okubo, T. Odagaki, J. Phys.: Condens. Matter {\bf 16}, 6651 (2004) 
\bibitem{Xuetal05} N. Xu, J. Blawzdziewicz, C.S. O’Hern, Phys. Rev. E {\bf 71}, 061306 (2005)
\bibitem{Koetal16} D. J. Koeze, D. V{\aa}gberg, B.B.T. Tjoa, B.P. Tighe, EPL {\bf 113}, 54001 (2016) 
\bibitem{Br25} H.J.H. Brouwers, arxiv: 2501.16005 (2025) and references therein
\bibitem{MaBl14} T. Matsushima, R. Blumenfeld, Phys. Rev. Lett. {\bf 112}, 098003 (2014)
\bibitem{Waetal20} C.C. Wanjura, P.A. Gago, T. Matsushima, R. Blumenfeld, Granular Matter {\bf 22}, 91 (2020)
\bibitem{SM} Supplemental material containing detailed calculations and more plots
\bibitem{Fe25} T. Fernique, Comp. Geom. Volumes {\bf 124–125}, 102134 (2025) 
\bibitem{SuZh26} S. Su and J. Zhang, Private communication (2026)
\bibitem{Fl60} A. Florian,  Rendiconti del Circolo Matematico di Palermo {\bf 9}, 300 (1960)
\bibitem{ZePr27} F. Zernike, J.A. Prins, Zeitschrift f\"ur Physik {\bf 41}, 184 (1927)
\bibitem{Stetal83} P.J. Steinhardt, D.R. Nelson, M. Ronchetti, Phys. Rev. B {\bf 28}, 784 (1983)
\bibitem{Jietal11} Y. Jiao;  F.H. Stillinger, S. Torquato, J. Appl. Phys. {\bf 109}, 013508 (2011)
\bibitem{Moetal24} P.K. Morse, P.J. Steinhardt, S. Torquato, Phys. Rev. Res. {\bf 6}, 033260 (2024)
\bibitem{MaBl17} T. Matsushima, R. Blumenfeld, Phys. Rev. E {\bf 95}, 032905 (2017)

\end{thebibliography}
\end{document}